\documentclass[twocolumn,showpacs,preprintnumbers,amsmath,amssymb,widetext]{revtex4}
\usepackage{psfig,graphics}
\usepackage[dvips]{graphicx}

\begin{document}
\title{High-performance planar light-emitting diode}
\author{Marco Cecchini}
\email{cecchini@sns.it}
\author{Vincenzo Piazza}
\author{Fabio Beltram}
\affiliation{NEST-INFM and Scuola Normale Superiore, I-56126 Pisa, Italy}
\author {Marco Lazzarino}
\affiliation{Laboratorio Nazionale TASC-INFM, Basovizza, I-34012 Trieste, Italy}
\author{M. B. Ward}
\author{A. J. Shields}
\affiliation{Toshiba Research Europe Ltd, 260, Science Park, Cambridge CB4 0WE, United Kingdom}
\author{H. E. Beere}
\author{D. A. Ritchie}
\affiliation{Cavendish Laboratory, University of Cambridge, Cambridge CB3 0HE, United Kingdom}

\begin{abstract}
Planar light-emitting diodes (LEDs) fabricated within a single high-mobility quantum well are demonstrated. Our approach leads to a dramatic reduction of radiative lifetime and junction 
area with respect to conventional vertical LEDs, promising very high-frequency device operation.
Devices were fabricated by UV lithography and wet chemical etching starting from p-type modulation-doped Al$_{0.5}$Ga$_{0.5}$As/GaAs heterostructures grown by molecular beam epitaxy. Electrical and optical measurements from room temperature down to 1.8\,K show high spectral purity and high external efficiency. Time-resolved measurements yielded extremely short recombination times of the order of 50\,ps, demonstrating the relevance 
of the present scheme for high-frequency device applications in the GHz range.
\end{abstract}

\maketitle

Widespread diffusion of optical communication systems
demands low-cost and high-performance light sources. High-radiance, wide-modulation-bandwidth light-emitting diodes (LEDs)
would be promising candidates to replace laser sources, especially for
short-haul in-fiber optic links. These sources offer many advantages thanks to the
simplicity of the device structure, ease of fabrication, high reliability and simplified biasing arrangement. Conventional vertical LEDs, however, suffer from intrinsic bandwidth and efficiency limitations
due to minority-carrier lifetime, internal capacitance,
and non-radiative recombination processes. 

Several fabrication schemes were proposed to improve
performance, mainly based on 
heavily-doped materials \cite{lyon,lyon2,chen}.
Introducing impurities strongly reduces the total recombination time
leading to increased modulation bandwidth, but it also affects
luminescence efficiency by introducing non-radiative recombination
channels.
The device
performance is thus a compromise
between modulation
bandwidth and radiative efficiency.
High-speed operation (in the GHz range) was reported with typical
efficiencies of few $\mu$W/mA \cite{lyon,lyon2,chen}.

Here we demonstrate a different approach based on the reduction of
junction-dimensionality and on a planar geometry. 

Our approach takes advantage of the extremely short recombination
times that can be obtained in a two-dimensional modulation-doped quantum well
(QW)\cite{sanvitto}. The doping scheme ensures  an increased radiative efficiency by drastically reducing the non-radiative recombination channels.
Additionally, our planar configuration leads to a drastic 
reduction of the parasitic capacitance. 
These aspects make
planar-junction devices very promising from the point of view of  
modulation bandwidth and spectral purity.
Furthermore, when an appropriate cavity is included in the design, our scheme may allow laser-emission by cold-electron injection into the active region. The planar configuration imposes no restrictions on cavity geometry, allowing the fabrication of both horizontal- and vertical-cavity devices.

The few existing realizations of pn junctions with lateral geometries \cite{saiki,vaccaro,north} are based on the
amphoteric nature of Si in GaAs. Under optimized growth conditions, Si
acts as an acceptor on the (311)A-oriented plane and as a donor on the
higher-index (n11)A-oriented planes. Therefore, it is possible to
obtain p- and n-type adjacent doped regions on a properly processed  
Si-doped GaAs substrate\cite{miller}. This approach
leads to a significant reduction of the junction area and to good
performance in terms of electro-luminescence, but the presence of
doping impurities in the conduction layers negatively impacts 
transport properties and the optical performance of the devices.

Our fabrication scheme is shown in the upper inset of
Fig.\,1(a). 
The starting system was a p-type modulation-doped Al$_{0.5}$Ga$_{0.5}$As/GaAs 
heterostructure with a two-dimensional hole gas (2DHG) confined 
within the GaAs quantum-well. This 2DHG constitutes the p-type portion of the final device. The heterostructure was processed into mesas 
and p-type Au/Zn/Au (5/50/150\,nm) Ohmic contacts were evaporated and 
annealed (60\,s at 460\,$^{\circ}$C) in nitrogen overpressure.

The fabrication of the n-type region involved two processing steps:
an ortophosphoric-acid-based etching solution
($\mathrm{H_{3}PO_{4}:H_{2}O_{2}:H_{2}O=3:1:50}$ for 95 seconds) was used to completely remove the Be-doped Al$_{0.5}$Ga$_{0.5}$As 
layer from a portion of the mesa to obtain an intrinsic region within the QW. Following the etching procedure, a self-aligned, n-type contact consisting of AuGe(eutectic)/Ni/Au (120/20/140\,nm) was evaporated. A 95\,s annealing at 450\,$^{\circ}$C in nitrogen overpressure completed the procedure.
The n-type contact introduces donors into the host semiconductor, creating an electron gas within the GaAs layer below the metal pad, adjacent to the 2DHG.

Several pn junctions were fabricated following the above procedure,
from a molecular-beam-epitaxy-grown heterostructure whose layer sequence is shown in the lower inset of Fig.\,1(a). Nominal Be concentration was 10$^{18}$\,cm$^{-3}$. Free-hole concentration in the 20\,nm-wide QW was numerically calculated by a Poisson-Schr\"odinger solver
which yielded a density of $8.8\times10^{10}$\,holes/cm$^{2}$.

Devices were tested by current-voltage (I-V) measurements from room temperature down to 1.8\,K. All fabricated devices presented very reproducible rectifying
characteristics, with threshold around
1.5\,V and negligible reverse-bias current ($< 20$\,nA at room temperature) in all the explored voltage range 
 (see Fig.\,1(a)). The diode-like behavior and
threshold values consistent with the GaAs energy-gap 
represent a first
evidence of the actual formation of the planar pn junction.

Spatially, spectrally and temporally-resolved
electro-luminescence was measured 
in a wide range of temperatures in order to characterize 
the optical properties of the devices. 
Light emission was obtained
from all the devices measured for forward biases above the turn-on value.
Data were taken from room temperature down to 1.8\,K.
The spatial distribution of the emitted light was measured
using an experimental set-up based on a low-vibration cold-finger cryostat.
The light-collection system consisted of a 100$\times$ objective coupled to a multi-mode optical fiber. The system was mounted on a computer-controlled motorized x-y translation stage to allow spatially-resolved measurements.
The signal was detected by a cooled photomultiplier after spectral filtering by a single-grating monochromator. Figure\,1(b) shows an electro-luminescence profile at room temperature under 6\,V forward bias superimposed onto the image of one device. The
darker region between the contacts is the mesa, from which emission
is observed. 
The spatial distribution of the emission was found to be markedly temperature
and drive current sensitive.  
At the lowest temperatures, the emission was found
to originate from around the p-type contact at moderately high biases, 
demonstrating that
the recombining electrons are able to travel a considerable distance before recombination.

Figure\,2 shows electro-luminescence spectra taken at
15\,K under a 2\,V forward bias. 
The emission peak is centered at 1522\,meV with a full width at half maximum (FWHM) of 7.4\,meV. Data show remarkably high spectral purity, much higher than available with conventional vertical or even with planar junctions obtained on patterned GaAs (311){\em A}-oriented substrates.

The observation of intense electro-luminescence and its peculiar spectral features demonstrates the functionality of the pn planar-junction scheme.

Emitted electroluminescence power was measured as a function of injected current and temperature (from 2.5\,K to room temperature). Assuming isotropic emission we were able to estimate device efficiency as the ratio of the integrated collected power over the injected current. Maximum efficiency was obtained at low temperatures and low injected currents (I $\simeq 0.187$\,mA) and was found to reach $\sim 13\,{\rm \mu W/mA}$.
By increasing the current, efficiency drops to about 1\,${\rm \mu
W/mA}$ (I $\simeq2$\,mA) and then remains nearly constant up to 8\,mA. 
At room temperature the current dependence of the emission efficiency  
is less pronounced and varies from 4.1\,${\rm \mu W/mA}$ to 2.5\,${\rm \mu W/mA}$ in the range of currents investigated.

The temporal evolution of the electroluminescence after short electrical excitation pulses was 
studied using a similar set-up consisting of a microscope objective, single grating spectrometer
and a Si avalanche photodiode detector. 
Time-resolved electroluminescence was measured
at 5\,K after excitation with 150\,ps voltage pulses.  An exponential
fit to the
curves yielded typical decay times
of around 300\,ps, close to the resolution limit of the avalanche photodiode detector.  Although
our set-up did not allow the temporal characteristic to be fully resolved, the measurements demonstrate 
that the emission can be modulated on sub-ns time scales.  
The measurements also showed that emission from spatial positions away from
the n-contact is delayed relative to emission next to the contact.  We
suppose this delay, which is of the order of a few nanoseconds, is due to
the propagation delay of electrons travelling along the mesa.

High-temporal-resolution photoluminescence measurements were also made
on the 2DHG using a Streak camera set-up.  For these measurements the luminescence was excited 
by 1\,ps optical pulses at a repetition rate of 80\,MHz and detected using a spectrometer and Streak 
camera in synchroscan mode.  The density of the hole gas was varied by applying a bias between an
Ohmic contact and a semi-transparent Schottky contact on the mesa surface.  
Figure\,3 shows how the photoluminescence lifetime varies with the applied Schottky gate bias.  
The
lifetime of the photo-excited carriers was found to be extremely 
short, ranging from
52\,ps to 242\,ps. Lifetime increases sharply as the QW is depleted. 
In the presence of a dilute density of excess holes, photo-induced formation of positively charged excitons (X$^{+}$) from free excitons (X) becomes
important\cite{shields}. X$^{+}$ excitons decay much more rapidly
because of the presence of a massive particle in the final state which
easily allows momentum conservation. 
By lowering the hole density, the formation of X$^{+}$ excitons is suppressed
resulting in an increase in the population of X.
In this regime, which we observed for positive gate voltages, decay of neutral 
excitons dominates radiative recombination and leads to the increased measured lifetime. These measurements demonstrate the high quality of
our QW structures and point out the suitability of this approach for the 
implementation of high-frequency devices in the GHz range.

In conclusion, we fabricated LEDs using a novel
planar-design scheme. Lateral junctions between n- and
p-type high-quality two-dimensional systems were fabricated starting from modulation-doped AlGaAs/GaAs heterostructures containing a 2DHG
within a GaAs QW. Devices were characterized in a wide range of
temperatures by current-voltage measurements, spatially and 
spectrally resolved electro-luminescence and 
emission-power measurements. Very good performance was obtained in terms of
spectral purity and high efficiency.
The small junction area and the extremely short minority-carrier lifetimes
make these devices ideally suitable for high-frequency operation.
\\

This work was supported in part by the European Commission through the FET Project
SAWPHOTON. M.C. acknowledges support by the C.N.R.


\begin{figure}[!ht]
\begin{center}
\includegraphics[width=0.4\textwidth]{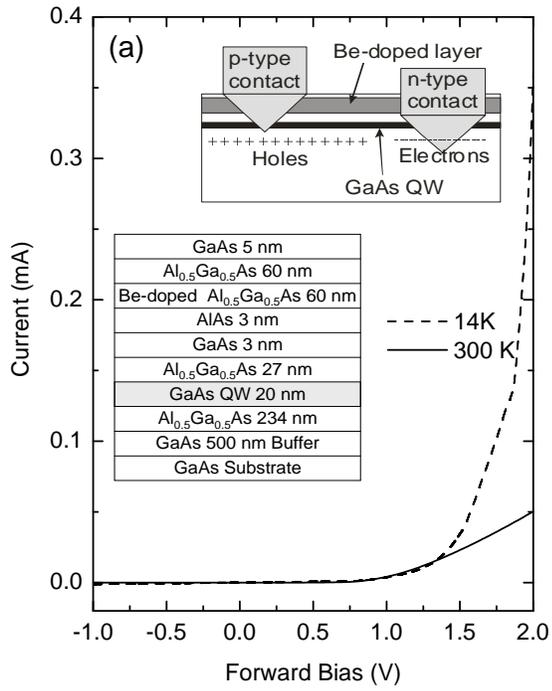}
\vspace{0.5cm}
\hspace{3cm}
\includegraphics[width=0.35\textwidth]{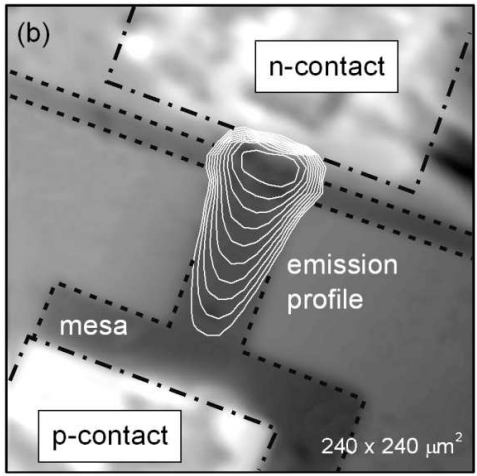}
\caption{(a) Current-voltage characteristic at T = 14\,K (dashed
line) and T = 300\,K (solid line) of the lateral pn-junction devices. Left inset: 
composition of the two heterostructures used. The right inset
reports the fabrication scheme of the devices. (b) Spatially and
spectrally resolved
electro-luminescence intensity profile (logarithmic scale) 
of a lateral pn-junction
at T = 300\,K for a forward voltage of 6\,V
superimposed onto a topology image of the device. The monochromator
was set at 860\,nm.}
\end{center}
\label{fig1}
\end{figure}

\begin{figure}[!ht]
\begin{center}
\includegraphics[angle=-90,width=0.4\textwidth]{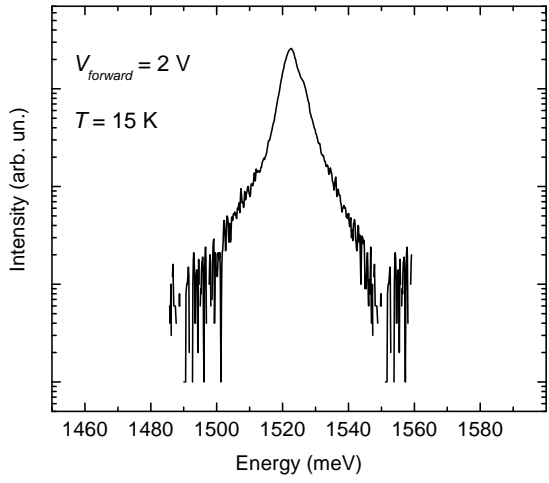}
\caption{Emission spectrum for a
forward bias of 2\,V at T = 15\,K. 
The intensity is plotted in logarithmic scale.} 
\label{fig2}
\end{center}
\end{figure}

\begin{figure}[!ht]
\begin{center}
\includegraphics[angle=-90,width=0.4\textwidth]{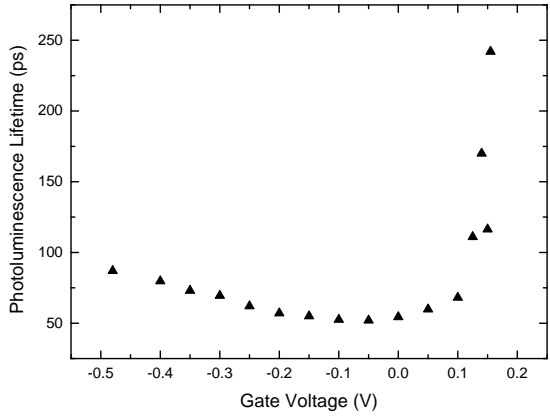}
\caption{Photoluminescence lifetime as a function of
gate voltage at 5\,K. The recombination processes are extremely
fast, with a minimum lifetime of 52\,ps.}
\end{center}
\label{fig3}
\end{figure}

\end{document}